\documentclass[twocolumn,showpacs,preprintnumbers]{revtex4}
\usepackage{amssymb}
\usepackage{amsfonts}
\usepackage{amsmath}
\usepackage{graphicx}
\usepackage{dcolumn}
\usepackage{bm}

\input{tcilatex}

\begin{document}

\title{Resistivity of two-dimensional systems in a magnetic field at the
filling factor $\nu =1/2$ }
\author{S. S. Murzin}
\affiliation{Institute of Solid State Physics RAS, 142432, Chernogolovka, Moscow
District, Russia}
\date{\today }

\begin{abstract}
Experimental data available in the literature on the diagonal resistivity $%
\rho _{xx}$ of GaAs/AlGaAs heterostructures in a magnetic field at the
filling factor $\nu =1/2$ have been compared with the existing theoretical
prediction [B. I. Halperin et al., Phys. Rev. B \textbf{47}, 7312 (1993) and
F. Evers et al., Phys. Rev. B \textbf{60}, 8951 (1999)]. It has been found
that the experimental results disagree with the prediction.
\end{abstract}

\maketitle

\begin{figure}[b]
\includegraphics[width=\columnwidth]{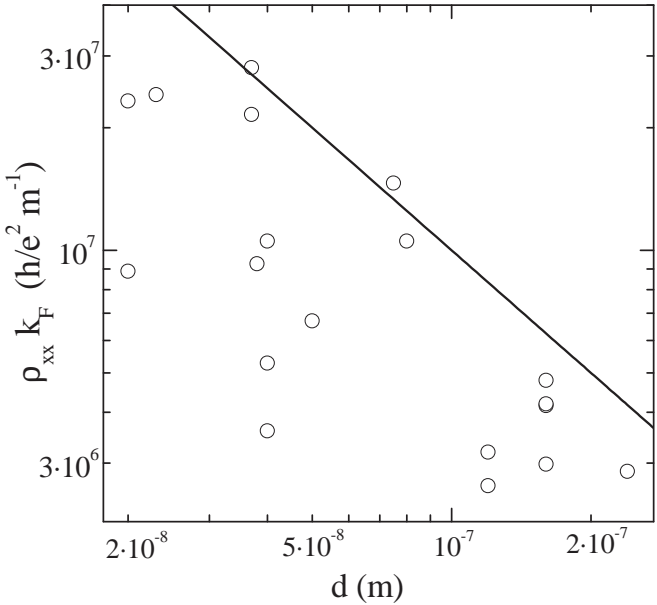}
\caption{$\protect\rho _{xx}(1/2)k_{F}$ versus the spacer thickness $d$. The
circles are the experimental data, and the straight line corresponds to
Eq.(2).}
\label{Rn1}
\end{figure}

\begin{figure}[b]
\includegraphics[width=\columnwidth,clip]{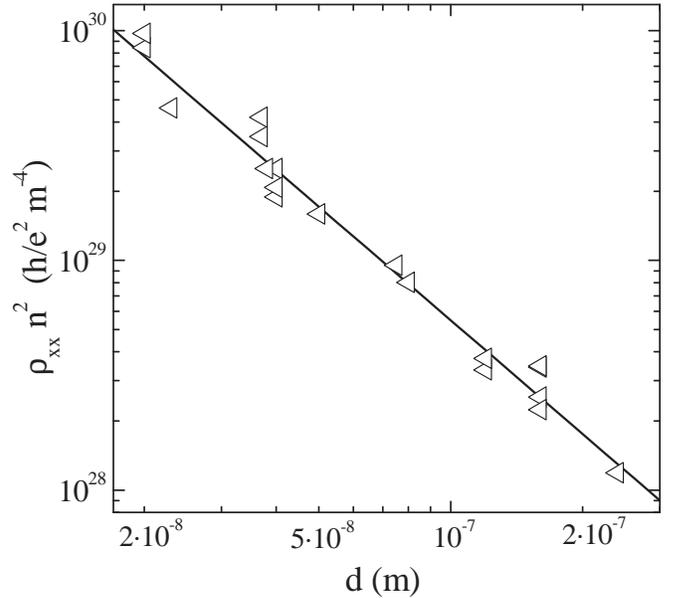}
\caption{$\protect\rho _{xx}n^{2}$ versus the spacer thickness $d$. The
triangles are the experimental data, and the solid line ($\protect\rho %
_{xx}n^{2}=1.79\times 10^{17}d^{-1.64}$) is the linear fit of the data in
the log--log scale.}
\label{Rn2}
\end{figure}

\begin{figure}[b]
\includegraphics[width=\columnwidth,clip]{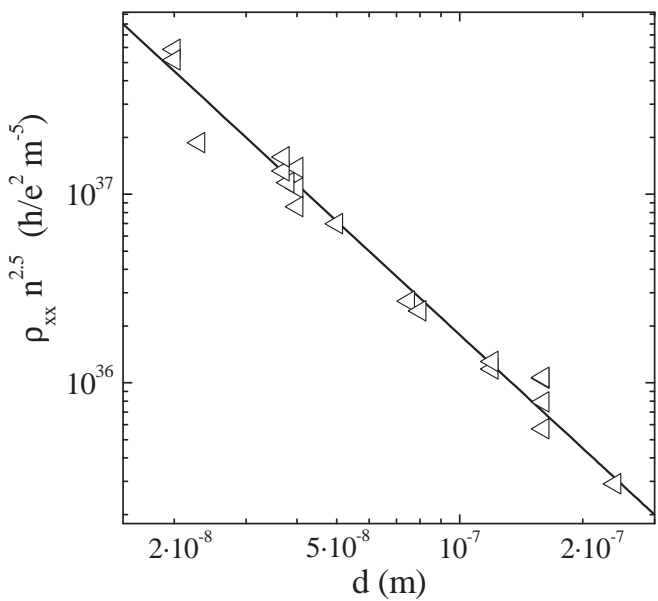}
\caption{$\protect\rho _{xx}n^{2.5}$ versus the spacer thickness $d$ The
triangles are the experimental data, and the solid line corresponds to
dependence $\protect\rho _{xx}n^{2.5}=$\ $1.8\times 10^{22}d^{-2}$ m$^{-3}$.}
\label{Rn3}
\end{figure}
\begin{figure}[b]
\includegraphics[width=\columnwidth,clip]{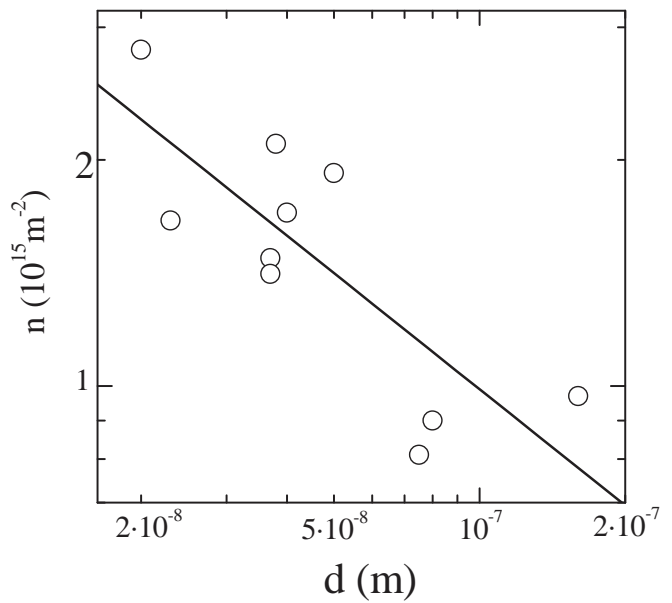}
\caption{Electron density $n$ in non-irradiated samples versus the spaces
thickness $d$ (circles). The line is a guide for the eyes.}
\label{nd}
\end{figure}

The expression for the diagonal resistivity $\rho _{xx}(1/2)$ of
two-dimensional systems in a magnetic field at the filling factor $\nu
=nh/(eB)=1/2$ was obtained in [1] based on the composite fermion theory. The
composite fermions are scattered by a random magnetic field induced by
ionized impurities. The ionized impurity concentration $n_{i}$ in the ideal
selectively-doped two-dimensional sample is equal to the electron density $n$
and, at $\nu =1/2$,

\begin{equation}
\rho _{xx}(1/2)\sim \frac{1}{k_{F}d}\frac{h}{e^{2}}.  \label{Rxx1}
\end{equation}%
Here $k_{F}=\sqrt{4\pi n}$, is the wavenumber of the composite fermions at
the Fermi level and $d$ is the spacer thickness. A more general and detailed
analysis of $\rho _{xx}(1/2)$ carried out in [2] yields the same but more
accurate result for the $n_{i}=n$ case:

\begin{equation}
\rho _{xx}(1/2)\approx 1.0\frac{1}{k_{F}d}\frac{h}{e^{2}}.  \label{Rxx2}
\end{equation}%
In this work, Eq.(2) is compared with the published experimental data
[3--17] on $\rho _{xx}(1/2)$ of single GaAs/AlGaAs heterojunctions with one
doped layer.

We used the $\rho _{xx}(1/2)$ data for the samples without gate, with the
mobility $40<\mu <900$~m$^{2}$/Vs, the spacer thickness $20\leq d\leq 240$%
~nm, and the electron density $6\times 10^{14}\leq n<5\times 10^{15}$~m$%
^{-2} $ at temperatures $0.047\div 1.3$ K. The data for samples with $n\leq
4.5\times 10^{14}$~m$^{-2}$ are not described well by presented below
expressions. The data for only two samples with $n\geq 6\times 10^{15}$~m$%
^{-2}$ were not used. The fractional quantum Hall effect in these samples
was developed substantially weaker than that in the other samples with the
close parameters. For the sample used in \cite{Kuk} whose resistance
depended on its prehistory, we used the data for the minimum-disorder case.
Some samples \cite{Nik88,Clark90,Nic94,Fox94,Rok95,Col95,Nic96} were
irradiated by light, the others \cite%
{Tsui83B2,Tsui83L,Men,Haug87,Clark88ss,Clark88L,Koch,Kuk} were not.

The experimental data are compared with Eq. (\ref{Rxx2}) in Fig. 1 where we
plotted the experimental \cite{c} and theoretical dependences of $\rho
_{xx}(1/2)k_{F}$ on $d$ (here and below $\rho _{xx}$ is in units of $h/e^{2}$%
.). The experimental points exhibit a large spread below the theoretical
line. This means that it is not background impurities that are the cause of
the discrepancy as they would raise results above the theoretical line. We
tried to plot the experimental values $\rho _{xx}(1/2)n^{p}$ with different
integer and half-integer $p$ versus the spacer thickness $d$ and to fit the
dependences by linear functions in the log--log scale (let us remind that $%
k_{F}=\sqrt{4\pi n}$). The best fit was obtained for $p=2$ and the exponent
of $d$ equal to $-1.64$ (see Fig. 2). This corresponds to the expression 
\begin{equation}
\rho _{xx}(1/2)=\alpha n^{-2}d^{-1.64},  \label{Rxx3}
\end{equation}%
where $\alpha =1.8\times 10^{17}$ m$^{-2.36}$. Equation 
\begin{equation}
\rho _{xx}(1/2)=\beta n^{-2.5}d^{-2},  \label{Rxx4}
\end{equation}%
with $\beta =1.8\times 10^{22}$ m$^{-3}$describes experimental data nearly
as well as expression (\ref{Rxx3}) (see Fig.3). The coefficients $\alpha $
and $\beta $\ are independent of the magnetic field, since at a given
filling factor $\nu =nh/eB=1/2$ the magnetic field is unambiguously related
to the electron density $n$. Fig.\ref{Rn2} and \ref{Rn3} extra indicates
that the large spread in the data points in Fig.\ref{Rn1} does not result
from the presence of unintentional impurities or defects in the samples. The
regular arrangement of the points in Figs.\ref{Rn2} and \ref{Rn3} calls for
a new explanation of the electron transport in the magnetic field at $\nu
=1/2$.

Note that on average for non-irradiated samples, $n$ decreases with an
increase in $d$ (see Fig. 4), but the relative spread of the data points in
Fig. 4 is larger than that in Figs.\ref{Rn2} and \ref{Rn3}.

I would like to thank P.T. Coleridge for his critical remark. This work is
supported by the Russian Foundation for Basic Research and INTAS.

\end{document}